\documentclass[reprint,aps]{revtex4-2}
\usepackage{amsfonts}
\usepackage{amssymb}
\usepackage{amsmath}
\usepackage{tabularx,epsfig,color,tabularx,epstopdf}
\usepackage{graphicx}
\usepackage{dcolumn}
\usepackage{bm}
\usepackage[center]{subfigure}
\usepackage{dcolumn}
\usepackage{bm}

\begin{document}

\title{Rogue waves and instability arising from long-wave-short-wave
resonance beyond the integrable regime}
\author{Wen-Rong Sun$^{1}$}
\email{Corresponding author: sunwenrong@ustb.edu.cn}
\author{Boris A. Malomed$^{2}$}
\author{Jin-Hua Li$^{1}$}
\affiliation{
$^{1}$ School of Mathematics and Physics, University of Science and Technology Beijing, Beijing 100083, China\\
$^{2}$Instituto de Alta Investigaci\'{o}n, Universidad de Tarapac\'{a},
Casilla 7D, Arica, Chile} 

\begin{abstract}
We consider instability and localized patterns arising from long wave-short
wave (LWSW) resonance in the non-integrable regime numerically.
We study the stability and instability of elliptic-function periodic waves
with respect to subharmonic perturbations, whose period is a
multiple of the period of the elliptic waves. We thus find the modulational instability (MI)
of the corresponding \textit{dnoidal waves}. Upon varying parameters of dnoidal waves,
spectrally unstable ones can be transformed into stable states via the Hamiltonian Hopf
bifurcation. For \textit{snoidal waves}, we find a transition of the dominant instability scenario between
the MI and instability with a bubble-like spectrum. For \textit{cnoidal} waves, we produce
three variants of the MI. Evolution of the unstable states is also
considered, leading to formation of rogue waves on top of the elliptic-wave
and continuous-wave backgrounds.
\end{abstract}

\maketitle


\preprint{APS/123-QED}






\section{ Introduction}

The resonance between long and short waves occurs if the group velocity of a
short (high-frequency) wave is equal to the phase velocity of a long
(low-frequency) wave. A general theory for the interaction between short and
long waves was developed in 1977 by Benney~\cite{lwsw1}. It has revealed a
variety of phenomena relevant to a broad class of physical problems. The
long-wave-short-wave (LWSW) resonance has been predicted in the context of
the interaction of capillary-gravity and long gravity waves~in hydrodynamics
\cite{l1}, as well as in the interaction of short-wave surface waves and
long-wave internal ones in the ocean \cite{Hashizume,Duchene,Sulem}. In
plasma physics, in the framework of the Zakharov's system \cite{lws1}, the
LWSW resonance pertains to Langmuir solitons moving with velocities close to
the speed of sound. In quasi-one-dimensional molecular crystals, it is the
resonance between exciton and phonon fields in the Davydov's model~\cite%
{add1}, and in nonlinear optics the LWSW resonance provides a mechanism for
the generation of terahertz modes from optical waves~\cite{lwsw2}.

It is well known that the integrable LWSW resonance system can be solved by
means of the inverse scattering technique~\cite{inte1,inte2}, producing
analytical solutions for solitons, breathers and rogue waves (RWs)~\cite%
{inte3,inte4,inte5,inte6,inte7,inte8}. However, the LWSW resonance system is
usually not integrable in realistic physical settings. For example,
Chowdhury and Tataronis~\cite{lwsw2} showed that LWSW resonance can be
achieved in a second-order nonlinear negative-refractive index medium if the
short waves are represented by the negative-index branch. The novelty of
that work is the introduction of the second-order nonlinearity for the
efficient resonant coupling, which is different from pervious works for slow
light, where the ponderomotive force alone gives rise to the local
nonlinearity~\cite{lwsw3, lwsw4}. With the term resulting from the cascaded $%
\chi ^{(2)}$ nonlinearity, the governing nonlinear system [see Eq.~(\ref%
{lwswm}) below] for the amplitudes of the high- and low-frequency wave
packets coupled by the LWSW resonance, is not integrable. There have been
relatively few studies of non-integrable LWSW resonance systems. In
particular, a systematic characterization of stability and instability of
periodic waves, and nonlinear dynamical evolution (including RWs) for the
non-integrable LWSW resonance system, are still lacking, to the best of our
knowledge.

Our objectives in this work are:

(A) For integrable systems, it is well known that modulational instability
(MI) of periodic solutions (including plane waves and solutions expressed in
\ terms of elliptic functions) can lead to the formation of localized
patterns, such as solitons, RWs and breathers~\cite{cjb1}-\cite{mmi20}. As
we obtain periodic traveling-wave solutions of the non-integrable LWSW
resonance system (in Section II), it is natural to inquire if this system
features MI of the periodic traveling waves and if RWs can be generated by
the MI. Because system~(\ref{lwswm}) considered here is not integrable, we
address the MI numerically and analyze outcomes of systematic direct
simulations.

(B) Recently, stability of periodic waves with respect to superharmonic
perturbations (i.e., periodic perturbations sharing the period with the
underlying waves, including higher-order harmonics) for the LWSW-resonance
(Benney) system was studied~\cite{lwsw5}. Rather than assuming that periodic
perturbations are of the superharmonic type, we consider arbitrary periods,
including \textit{subharmonic} perturbations, with multiple periods, with
respect to the underlying waves. The extension of the analysis beyond the
superharmonic perturbations to the subharmonic ones is essential as there
are elliptic-function solutions which are stable with respect to
superharmonic perturbations but, nevertheless, are subject to MI. As shown
below in Section IV B, this problem can be addressed by introducing the
Floquet exponent $\mu $ and subharmonic perturbations. It is relevant to
stress that subharmonic perturbations have a wider physical relevance than
superharmonic ones, as one usually considers domains which are larger than
the period of the unperturbed solution (e.g., in the context of the ocean
wave dynamics~\cite{oc1,oc2}). We address the stability against the
subharmonic perturbations numerically, using the Hill's technique~\cite{hill}%
.

The rest of the paper is organized as follows. Three types of periodic
traveling-wave solutions of the non-integrable LWSW resonance system are
obtained in Section II. The linear stability problem in the framework of the
Floquet-Fourier-Hill theory is formulated in Section III. The stability and
instability of elliptic-function waves with respect to the subharmonic
perturbations are examined in Section IV. Spatiotemporal RW structures in
the present system, are studied numerically in Section V. Section VI
summarizes the findings.

\section{Periodic traveling-wave solutions}

The non-integrable LWSW resonance system for a complex-valued short -wave
envelope $S$ and a real-valued long-wave field $D$ is given by~\cite{lwsw2}
\begin{equation}
iS_{\hat{t}}+\lambda _{1}S_{xx}+\beta _{1}|S|^{2}S=\gamma _{1}DS,\quad D_{%
\hat{t}}=\mu _{1}\left( |S|^{2}\right) _{x},  \label{lwswm}
\end{equation}%
where real parameters $\lambda _{1}$, $\beta _{1}$, $\gamma _{1}$ and $\mu
_{1}$ measure effects of dispersion, nonlinearity, nonlinear coupling, and
radiation stress of the short waves acting on the long waves, respectively.
The meaning of fields $S$ and $D$ depends on the
particular physical context. In particular, system~(\ref{lwswm}) models the
resonant interaction of short and long water waves. Note that system~(\ref%
{lwswm}) is a reduction of the bidirectional Zakharov's system to the
unidirectional propagation~\cite{lws1}. The same LWSW-resonance system
describes the generation of terahertz fields from optical waves. I this
case, fields $S$ and $D$ denote the optical-wave envelope and the terahertz
wave, respectively~\cite{lwsw2}. When $\beta _{1}=0$, system~(\ref{lwswm})
is integrable~\cite{inte2}. In this paper, we consider the non-integrable
system ~(\ref{lwswm}) with $\beta _{1}\neq 0$. System~(\ref{lwswm}) can be
cast in a more convenient form
\begin{equation}
iS_{t}-S_{xx}+\beta |S|^{2}S=LS,\quad L_{t}=-\sigma \left( |S|^{2}\right)
_{x},  \label{lwswm1}
\end{equation}%
by means of rescaling
\begin{equation}
t=-\lambda _{1}\hat{t},\quad L=-\frac{\gamma _{1}}{\lambda _{1}}D,\quad
\beta =-\frac{\beta _{1}}{\lambda _{1}},\quad \sigma =-\frac{\mu _{1}\gamma
_{1}}{\lambda _{1}^{2}}.  \label{sigma}
\end{equation}

Three species of elliptic-function solutions can be obtained. Note that
\textit{snoidal} and \textit{dnoidal} solutions, based on elliptic functions
of the $\mathrm{sn}$ and $\mathrm{dn}$ types were reported in Ref. \cite%
{lwsw5}, while cnoidal-wave solutions, based on $\mathrm{cn}$ functions,
were absent. Defining
\begin{equation}
S(x,t)=e^{-i\omega t}\hat{S}(y,t),~\mathrm{with}~~y=x-ct,  \label{c}
\end{equation}%
~Eq. (\ref{lwswm1}) is written as
\begin{subequations}
\label{lwm2}
\begin{eqnarray}
&&i\hat{S}_{t}-ci\hat{S}_{y}+\omega \hat{S}-S_{yy}+\beta |\hat{S}|^{2}\hat{S}%
=L\hat{S}, \\
&&L_{t}-cL_{y}=-\sigma \left( |\hat{S}|^{2}\right) _{y}.
\end{eqnarray}%
Further, letting $\hat{S}=e^{-icy/2}\phi (y)$ and $L=\psi (y)$ and
integrating, we obtain
\end{subequations}
\begin{subequations}
\label{lwm3}
\begin{eqnarray}
&&\psi =\frac{\sigma }{c}\phi ^{2}+\gamma , \\
\left(\frac{d\phi}{dy}\right) ^{2} &=&\frac{A}{2}\phi ^{4}+B\phi ^{2}+2H,
\end{eqnarray}%
where $A\equiv \beta -\frac{\sigma }{c}$, $B\equiv \omega -\gamma -\frac{%
c^{2}}{4}$, $\gamma $ and $H$ being constants of integration.

Then, three types of the elliptic-function solutions are written as:

$\bullet $ Dnoidal waves $(\beta -\frac{\sigma }{c}<0)$:
\end{subequations}
\begin{equation}
\phi =A_{3}\mathrm{dn}(my,k),  \label{sn12}
\end{equation}%
where $A_{3}^{2}\equiv -\frac{2cm^{2}}{c\beta -\sigma }$, $\gamma \equiv
\frac{1}{4}(-c^{2}-8m^{2}+4k^{2}m^{2}+4\omega )$, and $H\equiv \frac{1}{2}%
A_{3}^{2}\left( k^{2}-1\right) m^{2}$.

$\bullet $ Snoidal waves $(\beta -\frac{\sigma }{c}>0)$:
\begin{equation}
\phi =A_{1}\mathrm{sn}(my,k),  \label{sn11}
\end{equation}%
where $A_{1}^{2}\equiv \frac{2ck^{2}m^{2}}{c\beta -\sigma }$, $\gamma \equiv
\frac{1}{4}(-c^{2}+4m^{2}+4k^{2}m^{2}+4\omega )$, and $H\equiv \frac{1}{2}%
m^{2}A_{1}^{2}$.

$\bullet $ Cnoidal waves $(\beta -\frac{\sigma }{c}<0)$:
\begin{equation}
\phi =A_{2}\mathrm{cn}(my,k),  \label{sn13}
\end{equation}%
where $A_{2}^{2}\equiv \frac{-2ck^{2}m^{2}}{c\beta -\sigma }$, $\gamma
\equiv \frac{1}{4}(-c^{2}+4m^{2}-8k^{2}m^{2}+4\omega )$ and $H\equiv -\frac{1%
}{2}A_{2}^{2}\left( k^{2}-1\right) m^{2}$.

Here $\phi $ are periodic functions with period $T=\frac{4K}{m}$ for
solutions~(\ref{sn11}) and~(\ref{sn13}), and $T=\frac{2K}{m}$ for solutions~(%
\ref{sn12}), where $K(k)$ denotes the complete elliptic integral of the
first kind, with $0\leq k<1$~\cite{efb}.

\section{The linearized problem in the framework of the Floquet-Fourier-Hill
theory}

To study the spectral stability of the elliptic-function solutions with
respect to $P$-subharmonic perturbations, where $P\ $is$~$an integer, we
consider
\begin{subequations}
\label{ps1}
\begin{eqnarray}
&&\hat{S}(y,t)=e^{-icy/2}\left[ \phi (y)+\epsilon u(y,t)+i\epsilon v(y,t)%
\right] , \\
&&L(y,t)=\psi (y)+\epsilon w(y,t),
\end{eqnarray}%
where $\epsilon $ is an infinitesimal coefficient, and $u$, $v$ and $w$ are
real functions. Substituting~expressions (\ref{ps1}) into Eq.~(\ref{lwm2})
and keeping the first-order terms in $\epsilon $ leads to
\end{subequations}
\begin{subequations}
\begin{eqnarray}
&&v\left( \frac{c^{2}}{4}-\beta \phi ^{2}+\psi -\omega \right) +v_{yy}=u_{t},
\\
&&-u_{yy}-\frac{1}{4}u\left( c^{2}-12\beta \phi ^{2}+4\psi -4\omega \right)
-w\phi =v_{t}, \\
&&-2\sigma \phi u_{y}-2\sigma u\phi ^{\prime }+cw_{y}=w_{t}.
\end{eqnarray}%
Looking for perturbation eigenmodes as
\end{subequations}
\begin{equation}
(u(y,t),v(y,t),w(y,t))=e^{\lambda t}(U(y),V(y),W(y))+\mathrm{c.c.}
\label{uvw}
\end{equation}
(where $\mathrm{c.c.}$ denotes the complex conjugate) yields the spectral
problem
\begin{equation}
\lambda \left(
\begin{array}{l}
U \\
V \\
W%
\end{array}%
\right) =\left(
\begin{array}{ccc}
0 & L_{1} & 0 \\
L_{2} & 0 & -\phi \\
L_{3} & 0 & c\partial _{y}%
\end{array}%
\right) \left(
\begin{array}{l}
U \\
V \\
W%
\end{array}%
\right) ,  \label{tmsp1}
\end{equation}%
where we the operators are%
\begin{eqnarray}
&&L_{1}\equiv \partial _{y}^{2}+\frac{c^{2}-4\omega -4\beta \phi ^{2}+4\psi
}{4}, \\
&&L_{2}\equiv -\partial _{y}^{2}-\frac{c^{2}-4\omega -12\beta \phi
^{2}+4\psi }{4}. \\
&&L_{3}\equiv -2\sigma \phi \partial _{x}-2\sigma \phi ^{\prime }.
\end{eqnarray}

As the coefficient functions of the linearized problem are periodic in $y$
with period $T$, we use the Fourier expansion for them, $\phi
^{2}=\sum_{n=-\infty }^{\infty }Q_{n}e^{i2n\pi y/T}$, $\psi =\sum_{n=-\infty
}^{\infty }R_{n}e^{i2n\pi y/T}$, $\phi =\sum_{n=-\infty }^{\infty
}S_{n}e^{i2n\pi y/T}$ and $\phi ^{\prime }=\sum_{n=-\infty }^{\infty
}F_{n}e^{i2n\pi y/T}$, where $Q_{n}$, $R_{n}$, $S_{n}$ and $F_{n}$ are the
respective Fourier coefficients. Further, the
periodicity of the coefficient functions of the spectral problem suggests to
decompose the perturbations using the Floquet theorem,
\begin{subequations}
\begin{eqnarray}
&&U(y)=e^{i\mu y}H_{U}(y)=e^{i\mu y}\sum_{n=-\infty }^{+\infty
}U_{n}e^{i2n\pi y/PT},  \notag \\
&&V(y)=e^{i\mu y}H_{V}(y)=e^{i\mu y}\sum_{n=-\infty }^{+\infty
}V_{n}e^{i2n\pi y/PT},  \notag \\
&&W(y)=e^{i\mu y}H_{W}(y)=e^{i\mu y}\sum_{n=-\infty }^{+\infty
}W_{n}e^{i2n\pi y/PT},  \notag
\end{eqnarray}%
where we expand $H_{U}(y)$, $H_{V}(y)$ and $H_{W}(y)$ as Fourier series in $y
$ with period $PT$, $\mu \in \lbrack 0,2\pi /T)$ is the Floquet exponent,
and
\end{subequations}
\begin{subequations}
\begin{eqnarray}
&&U_{n}\equiv \frac{1}{PT}\int_{-PT/2}^{+PT/2}H_{U}(y)e^{-i2\pi ny/PT}dy,
\notag \\
&&V_{n}\equiv \frac{1}{PT}\int_{-PT/2}^{+T/2}H_{V}(y)e^{-i2\pi ny/PT}dy,
\notag \\
&&W_{n}\equiv \frac{1}{PT}\int_{-PT/2}^{+PT/2}H_{W}(y)e^{-i2\pi ny/PT}dy.
\notag
\end{eqnarray}%
Substituting the Fourier expansions in Eq.~(\ref{tmsp1}) and equating the
Fourier coefficients results in the following bi-infinite spectral problem:
\end{subequations}
\begin{subequations}
\label{st}
\begin{eqnarray}
&&\left(- \omega +\frac{c^{2}}{4}+\left( i\mu +\frac{2in\pi }{PL}\right)
^{2}\right) V_{n}-\beta \sum_{m=-\infty }^{+\infty }Q_{\frac{n-m}{P}}V_{m}
\notag \\
&&+\sum_{m=-\infty }^{\infty }R_{\frac{n-m}{P}}V_{m}=\lambda U_{n}, \\
&&\left( \omega -\frac{c^{2}}{4}-\left( i\mu +\frac{2in\pi }{PL}\right)
^{2}\right) U_{n}+3\beta \sum_{m=-\infty }^{+\infty }Q_{\frac{n-m}{P}}U_{m}
\notag \\
&&-\sum_{m=-\infty }^{\infty }R_{\frac{n-m}{P}}U_{m}-\sum_{m=-\infty
}^{+\infty }S_{\frac{n-m}{P}}W_{m}=\lambda V_{n}, \\
&&-2\sigma \left( i\mu +\frac{2in\pi }{PL}\right) \sum_{m=-\infty }^{+\infty
}S_{\frac{n-m}{P}}U_{m}-2\sigma \sum_{m=-\infty }^{+\infty }F_{\frac{n-m}{P}%
}U_{m}  \notag \\
&&+c\left( i\mu +\frac{2in\pi }{PL}\right) W_{n}=\lambda W_{n}
\end{eqnarray}%
where $Q_{\frac{n-m}{P}},R_{\frac{n-m}{P}},S_{\frac{n-m}{P}}\quad \text{and}%
\quad F_{\frac{n-m}{P}}=0$ if $n-m$ is not divisible by $P$.
The spectral problem~(\ref{st}) is tantamount to~that based on Eq. (\ref%
{tmsp1}), and the spectrum of~(\ref{tmsp1}) is constructed as the union of
the spectra for all values of $\mu $.

\section{(In)stability of the elliptic-function waves}

Truncating the number of modes in the Fourier decompositions in the
bi-infinite spectral problem~(\ref{st}) to finite $N$, we calculate the
spectrum of the linear problem~(\ref{tmsp1}) numerically. The instability
growth rate, if any, is determined by a positive real part of $\lambda $.

\subsection{The modulational instability (MI) of the superharmonically
stable dnoidal waves}

For dnoidal waves (\ref{sn12}), Fig.~\ref{Gfig3} shows the largest
instability growth rate $\gamma $ as a function of $\beta $ and $c$ [see
Eqs. (\ref{c}) and (\ref{sigma})], while the other parameters are fixed. It
can be seen that simultaneously increasing $\beta $ and $|c|$ is favorable
for the stability, while simultaneously increasing $|c|$ and decreasing $%
\beta $ leads to the most unstable case.

The dnoidal waves are stable with respect to the superharmonic perturbations
(corresponding to $P=1$), as shown in the left panel of Fig.~\ref{Gfig4}.
Rather than assuming that perturbations share the period with the
unperturbed waves (thus restricting the consideration to the superharmonic
perturbations), we consider arbitrary periods, including subharmonic
perturbations. We find that, although the dnoidal waves are stable with
respect to the superharmonic perturbations, such waves are modulational
unstable with respect to the subharmonic perturbations, with the instability
band which has the shape of the figure-of-eight, (as shown in the middle
panel of Fig.~\ref{Gfig4}. Besides that, we note that, with the increase of $%
|c|$, the figure-of-eight-shaped instability band disappears, the dnoidal
waves being stable against all subharmonic perturbations (as shown in the
right panel of Fig.~\ref{Gfig4}). This implies that spectrally unstable and
stable states can convert into each other.

Next, we examine the transition from a spectrally unstable (stable) state to
a spectrally stable (unstable) one. We take $P=3$ as an example. When $%
c<c_{p}\approx -0.4118$, two eigenvalues (corresponding to $P=3$) are found
on the imaginary axis, as shown in the right panel of Fig.~\ref{Gfig5}. At $%
c>c_{p}$, the instability occurs when two critical imaginary eigenvalues
collide along the imaginary axis (as shown in the middle panel of Fig.~\ref%
{Gfig5}) through a Hamiltonian Hopf bifurcation~\cite{hhb}, and then enter the right
and left half planes along the figure-of-eight path, as shown in the left
panel of Fig.~\ref{Gfig5}).

Due to the presence of the subharmonic MI for dnoidal waves, we expect that
RWs may emerge on top of the dnoidal-wave background. These results are
reported in Section V.

\begin{figure}[tbph]
\centering
\includegraphics[height=150pt,width=200pt]{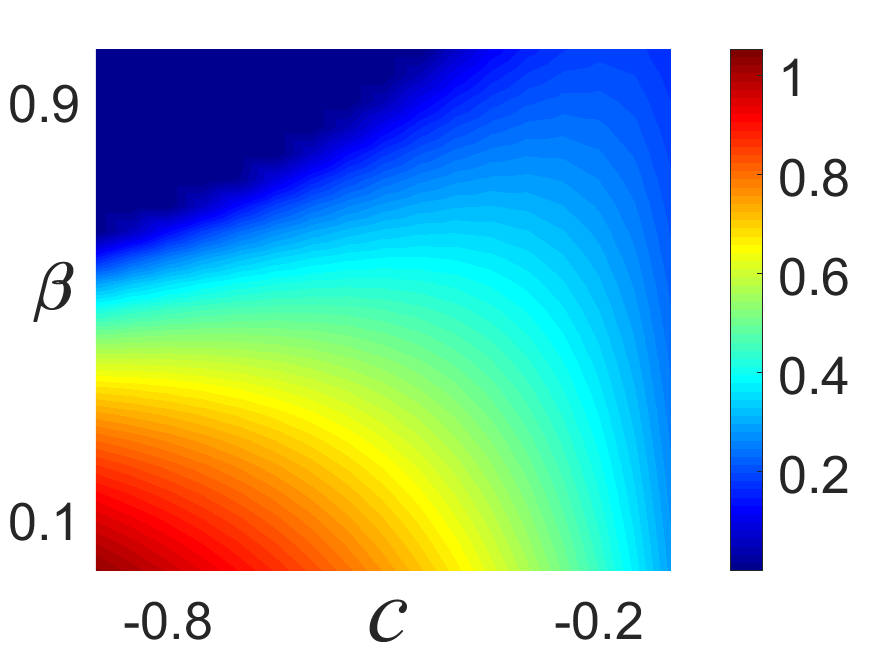} \newline
\caption{A color map of the largest instability growth rate $\protect\gamma $
as a function of $c$ and $\protect\beta $ for the dnoidal waves with $m=1$, $%
\protect\omega =-1$, $k=0.1$ and $\protect\sigma =-1$. }
\label{Gfig3}
\end{figure}

\begin{figure*}
\centering
\includegraphics[height=130pt,width=400pt]{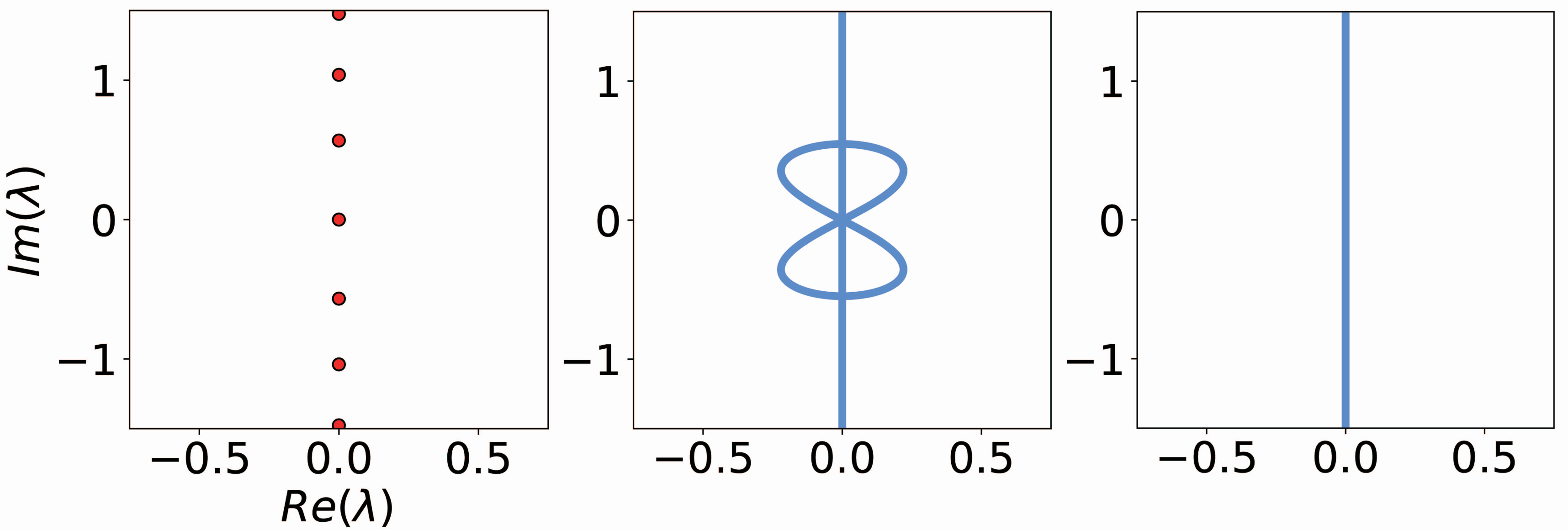} \newline
\caption{The spectrum of pertrubation eigevalues for the dnoidal waves with $%
\protect\omega =-1$, $m=1$, $k=0.1$, $\protect\sigma =-1$, $\protect\beta =0.9
$. The left panel shows the spectrum with respect to superharmonic
perturbations ($P=1$) for the dnoidal wave with $c=-0.2$ and the red dots represent the eigenvalues with $P=1$. The middle panel
shows the spectrum with respect to subharmonic perturbations for the wave
with $c=-0.2$. The right panel shows the spectrum with respect to
subharmonic perturbations for the wave with $c=-0.54$.}
\label{Gfig4}
\end{figure*}

\begin{figure*}
\centering
\includegraphics[height=130pt,width=422pt]{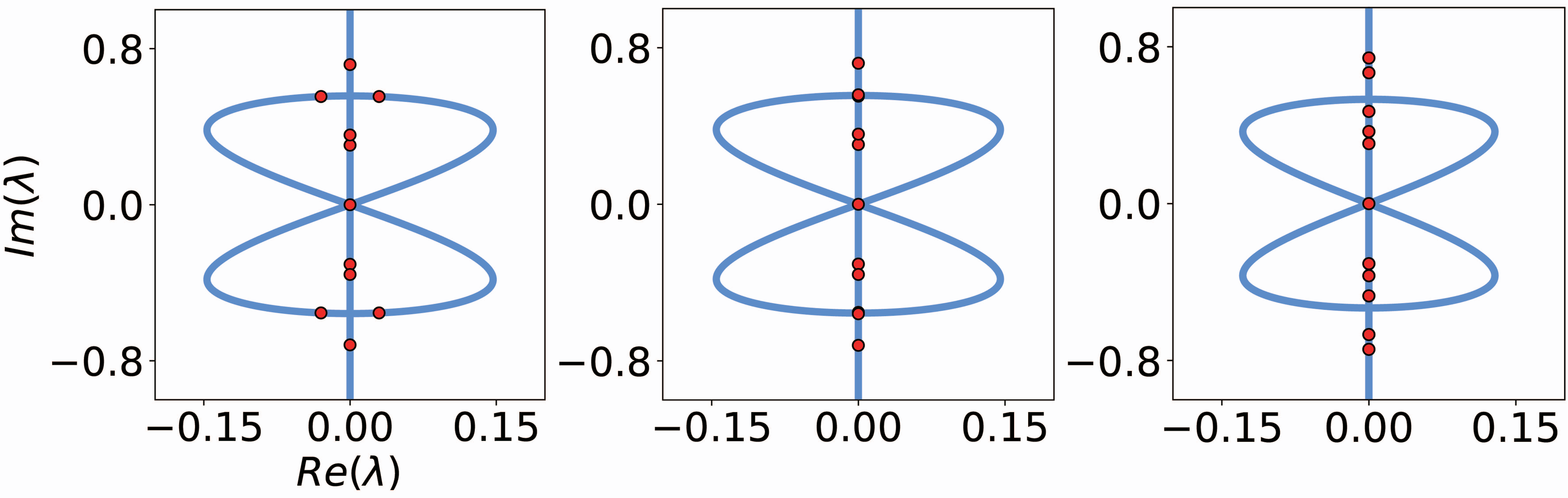}\newline
\caption{The Hamiltonian Hopf bifurcation: the perturbation spectrum for the
dnoidal waves with $\protect\omega =-1$, $m=1$, $k=0.1$, $\protect\sigma =-1$%
, $\protect\beta =0.9$, $c=-0.41$ (left), $c=-0.4118$ (middle), and $c=-0.43$
(right). Red dots represent the eigenvalues with $P=3$.}
\label{Gfig5}
\end{figure*}

\subsection{Instability transitions for snoidal waves and the modulational
instability (MI) for cnoidal waves}

To study instability-driven transitions for snoidal waves, at different
values of $c$, we consider the instability of these waves with respect to
subharmonic perturbations. Figure~\ref{Gfig1} shows the largest instability
growth rate $\gamma $ as a function of $c$. It is seen that $\gamma $
increases as $c$ increases. Figure~\ref{Gfig2} shows two types of the
instability: a bubble-like scenario (with the unstable band of the
eigenvalues shaped as a bubble-like curve), and MI (recall that MI involves
an unstable band of eigenvalues encompassing the origin). At $c<c_{\mathrm{cp%
}}$, where $c_{\mathrm{cp}}$ is a critical point for the instability
transition, MI is the dominant instability, as shown in the left panel of
Fig.~\ref{Gfig2}). As $c$ increases towards $c_{\mathrm{cp}}$, the MI band
is compressed horizontally (as shown in the middle panel of Fig.~\ref{Gfig2}%
), and the collision of the eigenvalues on the imaginary axis gives rise to
a large bubble-shaped  instability band and a figure-of-eight-shaped MI
band, as shown in the right panel of Fig.~\ref{Gfig2}). Note that the
maximal instability growth rate of that bubble-like instability is larger
than the maximal instability growth rate of that MI. This evolution implies
that the dominant instability switches into the bubble-like scenario.

\begin{figure}[tbp]
\centering
\includegraphics[height=138pt,width=140pt]{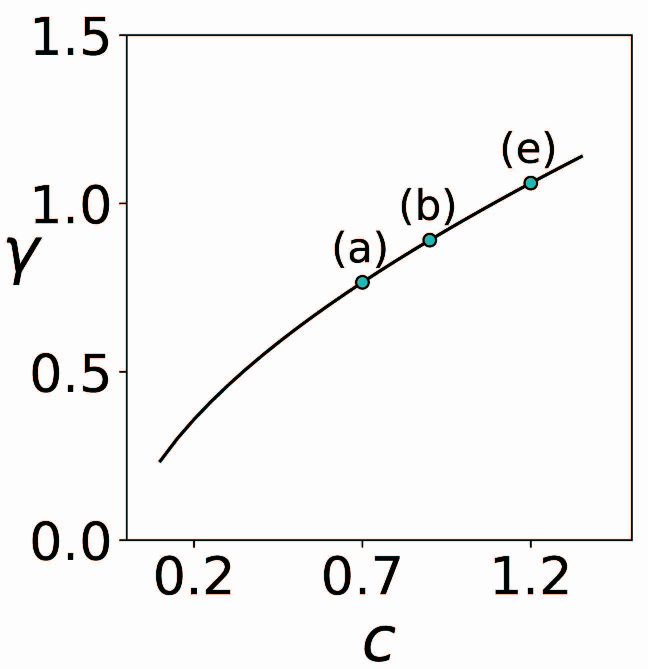} \newline
\caption{The largest instability growth rate $\protect\gamma $ as a function
of $c$ for snoidal waves with $\protect\beta =0.2$, $m=1$, $\protect\omega =-1
$, $k=0.8$, $\protect\sigma =-1$. Spectra of the perturbation eigenvalues
are shown in Fig. \protect\ref{Gfig2} for $c=0.7$ (a), $c=0.9$ (b), and $%
c=1.2$ (e).}
\label{Gfig1}
\end{figure}

\begin{figure*}
\centering
\hspace{-1cm}\includegraphics[height=105pt,width=185pt]{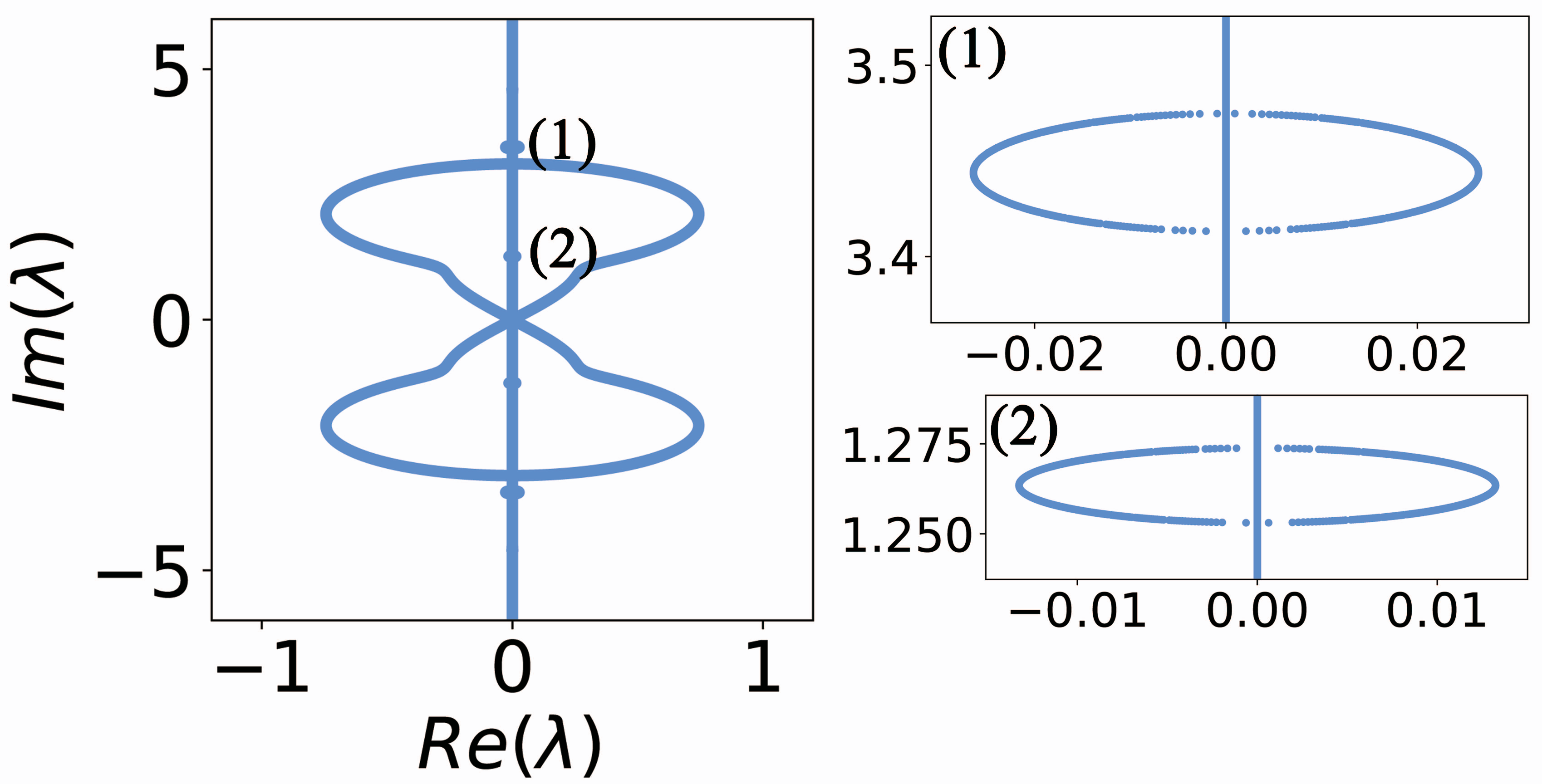}\includegraphics[height=105pt,width=185pt]{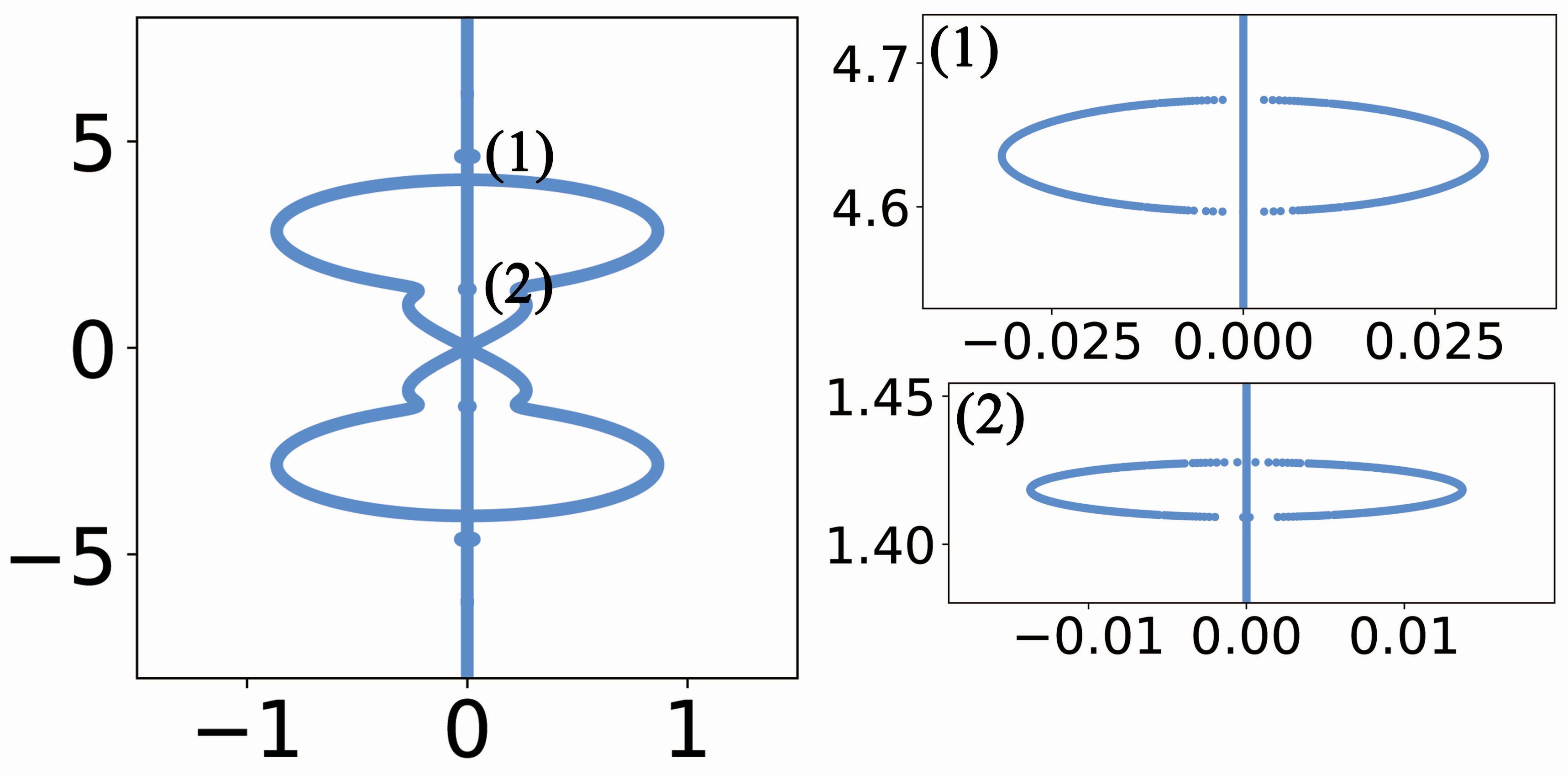}\includegraphics[height=105pt,width=170pt]{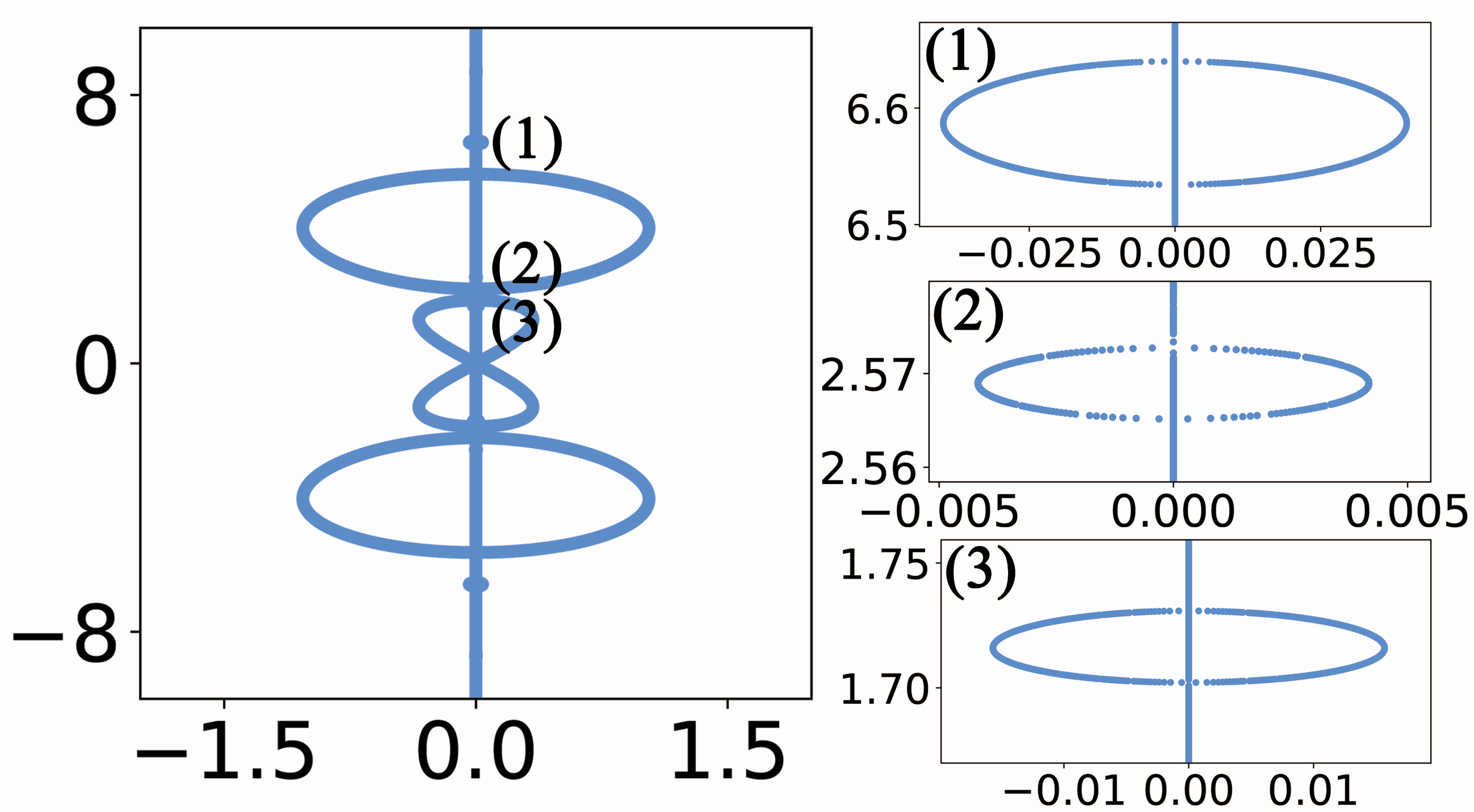}\newline
\caption{The spectrum of the subharmonic perturbations for snoidal waves,
with the parameters corresponding to Fig. \protect\ref{Gfig1} and (a) $c=0.7$%
, (b) $c=0.9$ and (e) $c=1.2$ (from left to right). }
\label{Gfig2}
\end{figure*}

\begin{figure*}
\centering
\includegraphics[height=130pt,width=400pt]{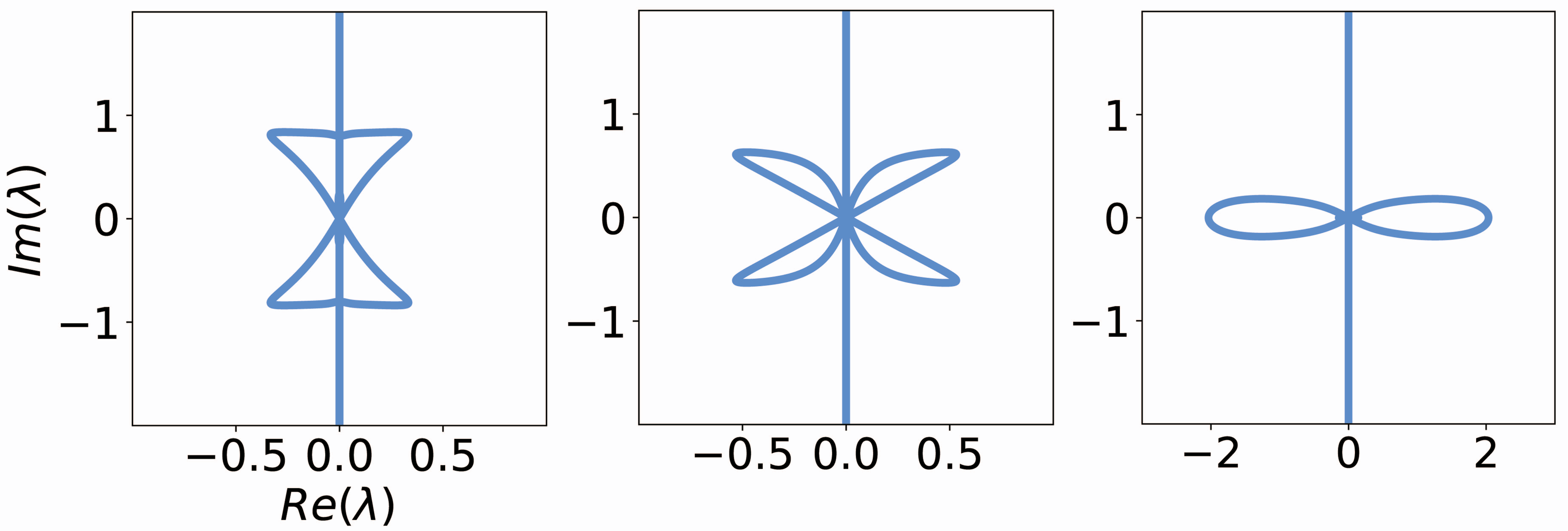} \newline
\caption{The perturbation spectrum (with respect to subharmonic perturbations) for cnoidal waves with $\protect\omega =-1$%
, $m=1$, $k=0.5$, $\protect\sigma =-1$, $c=-0.8$, and $\protect\beta =0.2$
(left), $\protect\beta =0.9$ (middle), and $\protect\beta =1.2$ (right).}
\end{figure*}
\label{Gfig6}

For the cnoidal waves, we show three variants of MI. When $\beta =0.2$, the
MI is shown in the left panel of Fig.~\ref{Gfig6}. As $\beta $ increases,
the instability band is pinched vertically (see the middle panel of Fig.~\ref%
{Gfig6}), as more eigenvalues from the imaginary axis accumulate in it. Then
the MI band features the shape of butterfly wings. Continuing to increasing
the value of $\beta $, the wings are pinched vertically, transforming the
shape of the MI band into an infinity symbol, as shown in the right panel of
Fig.~\ref{Gfig6}.

\section{Rogue waves in the non-integrable LWSW-resonance system}

The  instability of the elliptic-wave solutions makes it natural to search
for RW solutions emerging on top of the respective unstable backgrounds.
First, we address this issue for a simple case, \textit{viz}., generating
RWs by MI of continuous-wave (CW) backgrounds. Then we demonstrate the
emergence of RWs on top of the dnoidal wave, as an example for the
elliptic-wave background.

\subsection{Rogue waves (RWs) on the CW backgrounds}

It is well known that RWs can be generated by MI of the baseband type~\cite%
{mmi12}. Very recently~\cite{mmi19}, it has been shown that in the case of
the baseband or zero-wavenumber-gain MI, the mechanism for the RW formation
works solely under a linear relation between the MI gain and a vanishingly
small wavenumber of the modulational perturbations. These results were
obtained in integrable systems. In the present context, it is natural to
inquire whether the non-integrable LWSW-resonance system gives rise to
similar RW patterns on top of the CW under.

We here show an explicit condition under which the MI gain and perturbation
wavenumber satisfy an asymptotically linear relation. We will numerically
demonstrate that the RW can be created if this relation holds. The CW
solution is
\end{subequations}
\begin{equation}
S(x,t)=ae^{i(k_{1}x+\omega _{1}t)},L(x,t)=b,\omega _{1}=a^{2}\beta
-b+k_{1}^{2},
\end{equation}%
where real parameters $a$ and $b$, $k_{1}$ and $\omega _{1}$ are amplitudes,
wave number and frequency, respectively.

The perturbed solutions are expressed as
\begin{subequations}
\label{pss1}
\begin{eqnarray}
&&\hspace{-0.4cm}S(x,t)=e^{i(k_{1}x+\omega _{1}t)}(a+\eta _{1}(t)e^{i\Omega
x}+\eta _{2}(t)e^{-i\Omega x}), \\
&&L(x,t)=b+g_{1}(t)e^{i\Omega x}+g_{1}^{\ast }(t)e^{-i\Omega x},
\end{eqnarray}%
where $\Omega $ is the perturbation wavenumber. Substituting~expressions (%
\ref{pss1}) into Eq.~(\ref{lwswm1}) leads to the linear system
\end{subequations}
\begin{equation}
\left(
\begin{array}{c}
\eta _{1t} \\
\eta _{2t}^{\ast } \\
g_{1t}%
\end{array}%
\right) =iM\left(
\begin{array}{c}
\eta _{1} \\
\eta _{2}^{\ast } \\
g_{1}%
\end{array}%
\right) =i\left(
\begin{array}{ccc}
\Delta _{1} & a^{2}\beta  & -a \\
-a^{2}\beta  & \Delta _{2} & a \\
-a\sigma \Omega  & -a\sigma \Omega  & 0%
\end{array}%
\right) \left(
\begin{array}{c}
\eta _{1} \\
\eta _{2}^{\ast } \\
g_{1}%
\end{array}%
\right) ,  \label{em1}
\end{equation}%
where $\Delta _{1}\equiv a^{2}\beta +\Omega (2k+\Omega )$ and $\Delta
_{2}\equiv -a^{2}\beta -\Omega (\Omega -2k)$. The eigenvalues, which are
obtained as roots of the characteristic polynomial of $M$, \textit{viz}., $%
-2a^{2}\beta \Lambda \Omega ^{2}-2a^{2}\sigma \Omega ^{3}+4k^{2}\Lambda
\Omega ^{2}+\Lambda ^{3}-4k\Lambda ^{2}\Omega -\Lambda \Omega ^{4}=0$, may
be either real or form complex-conjugate pairs.

MI occurs if $\Lambda $ has a negative imaginary part, i.e., $\mathrm{Im}%
(\Lambda )<0$. This happens when the discriminant is negative, i,e.
\begin{subequations}
\begin{eqnarray}
&&4\Omega ^{6}\left( 8a^{6}\beta ^{3}-a^{4}\left( -12\beta ^{2}\Omega
^{2}+32\beta ^{2}k^{2}+72\beta k\sigma +27\sigma ^{2}\right) \right.   \notag
\\
&&\left. +2a^{2}\left( 3\beta \Omega ^{4}-16\beta k^{2}\Omega ^{2}+16\beta
k^{4}+8k^{3}\sigma -18k\sigma \Omega ^{2}\right) \right.   \notag \\
&&\left. +\left( \Omega ^{3}-4k^{2}\Omega \right) ^{2}\right) <0.
\end{eqnarray}

As we focus on the relation between the MI gain and a vanishingly small
wavenumber of the modulational perturbation, by considering $\Omega\rightarrow {0}$
 and $\Lambda =\Omega \hat{\Lambda}$, we write the characteristic
polynomial of $M$ as $-\hat{\Lambda}\left( 2a^{2}\beta +\Omega ^{2}\right)
-2a^{2}\sigma +4k^{2}\hat{\Lambda}-4k\hat{\Lambda}^{2}+\hat{\Lambda}^{3}=0$,
reducing the discriminant to $\Delta _{3}=4a^{2}\left( 8a^{4}\beta
^{3}-32a^{2}\beta ^{2}k^{2}-72a^{2}\beta k\sigma -27a^{2}\sigma ^{2}+32\beta
k^{4}+16k^{3}\sigma \right) $. Therefore, if $\Delta _{3}<0$, a linear
relation between the MI gain and a vanishingly small wavenumber of the
modulational perturbation is maintained. Then we expect that $\Delta _{3}<0$
may lead to the RW formation.

To check this prediction, we simulated the evolution of the CW states taken
as the initial condition, perturbed by a random Gaussian noise of relative
strength $5\%$. The domain used in the simulations is the same as in the pictures shown here.
The numerical computations were performed with fixed periodic boundary conditions
(matched to the constant plane-wave background).
As demonstrated in Figs.~\ref{fig1} and~\ref{fig2},
the noisy background features apparent MI-driven chaotic dynamics. For the
parameters which satisfy the above-mentioned condition $\Delta _{3}<0$, the
simulations reveal the formation of multiple isolated peaks that emerge at
random positions, which may be interpreted as the RWs.
In particular, the effects of $\beta$ on the generation of the first vector RWs on top of
the plane wave background are summarized in Table I. The time of
the emergence of the first RW component
$S$ ($L$) is denoted as $t_{S0}$ ($t_{L0}$). The increase of $\beta$ leads to the later
appearance of the first vector RW and reduction of its amplitude..

\begin{figure}[tbp]
\centering
\includegraphics[height=100pt,width=120pt]{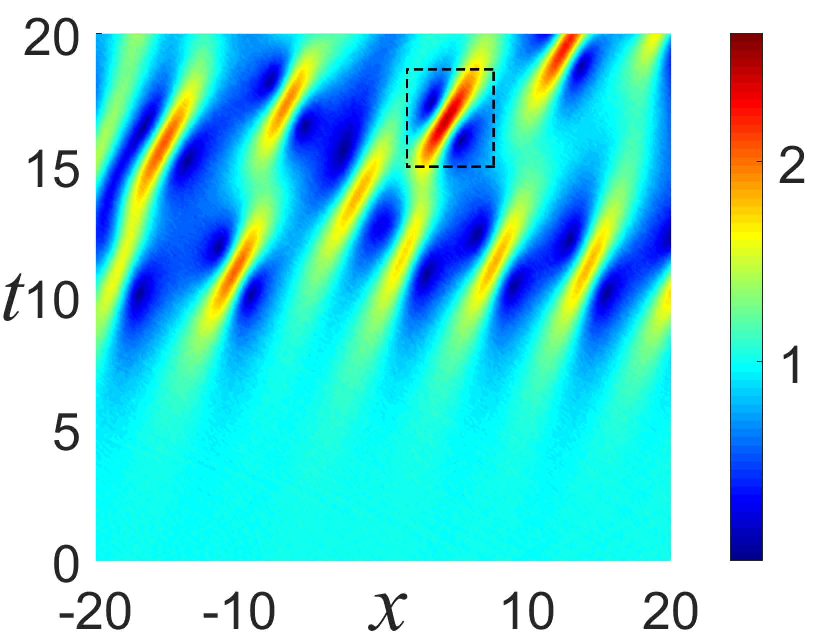} %
\includegraphics[height=100pt,width=120pt]{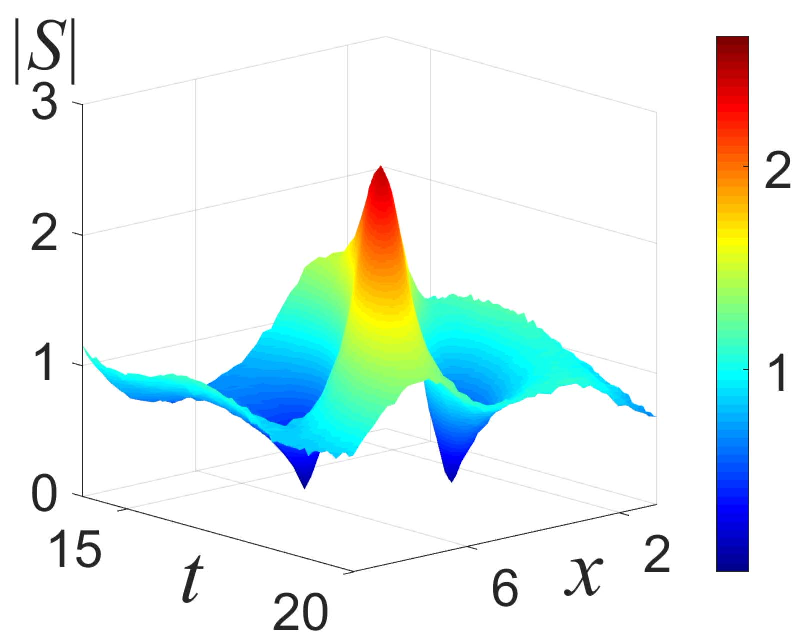} \newline
\caption{Numerically simulated formation of RWs, arising from the MI of the
CW background of the $S$-component of the non-integrable LWSW-resonance
system with $a=1$, $k=0$, $\protect\beta =0.1$, $\protect\sigma =0.5$ and $%
b=0.7$. A particular RW produced by the MI evolution is isolated by the
surrounding box. The right panel displays the three-dimensional zoom of this
RW.}
\label{fig1}
\end{figure}

\begin{figure}[tbp]
\centering
\includegraphics[height=100pt,width=120pt]{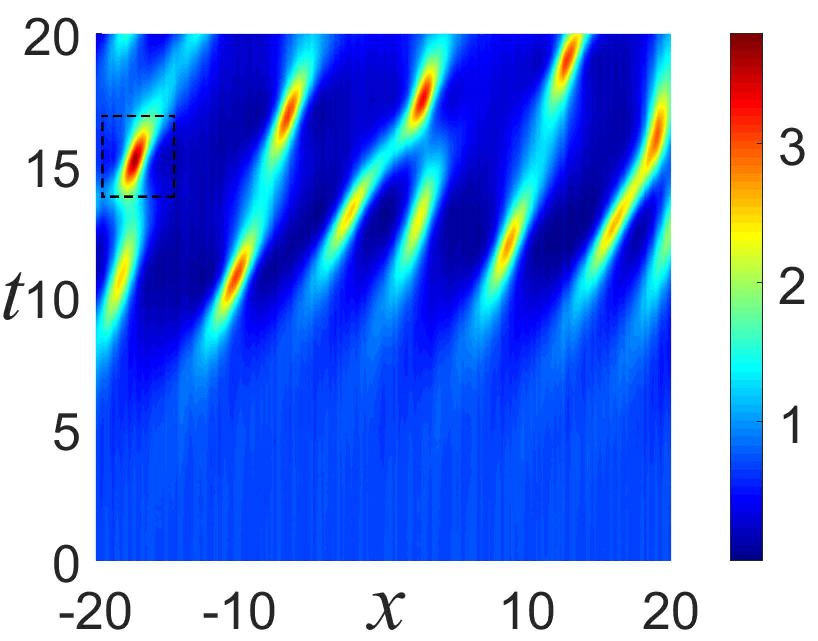} %
\includegraphics[height=100pt,width=120pt]{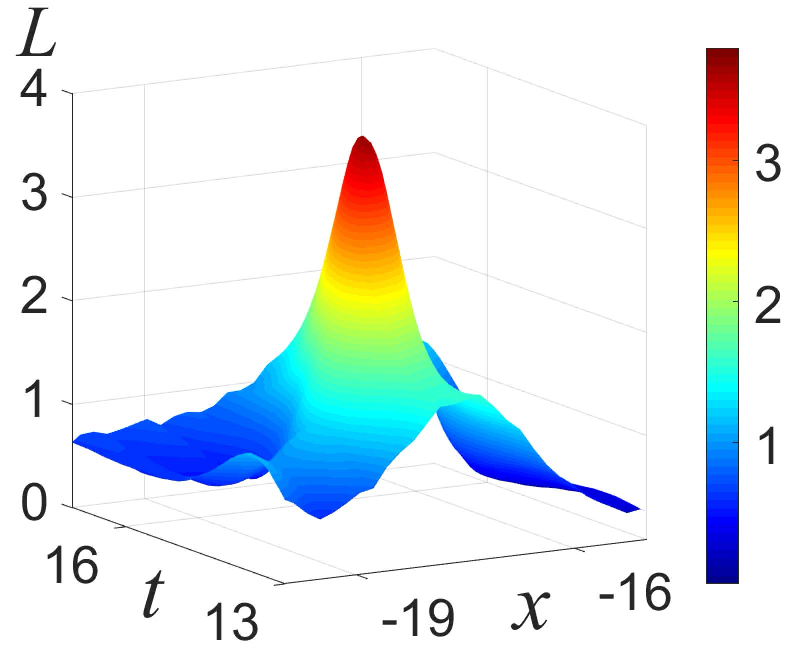} \newline
\caption{Numerically simulated formation of RWs, arising from the MI of the
CW background of the $L$-component of the LWSW-resonance system with $a=1$, $%
k=0$, $\protect\beta =0.1$, $\protect\sigma =0.5$ and $b=0.7$. A particular
emerging SW is isolated by the surrounding box. The right panel displays the
three-dimensional zoom of this RW.}
\label{fig2}
\end{figure}

\subsection{Rogue waves (RWs) on top of the elliptic-wave background}

Dynamics of rogue waves on the spatially-periodic background has attracted a significant research interest in certain integrable systems, such as the focusing nonlinear Schr\"{o}dinger equation~\cite{new1,new2,new3}. For example, the computation results of rogue waves on the spatially-periodic background were obtained firstly in~\cite{new1} with the help of numerical method and Darboux transformation. 
Then in~\cite{new2}, the authors obtained the exact analytical solutions for the rogue waves on the periodic background by computing exactly the branch points in the band-gap spectrum of the Zakharov-Shabat problem associated with the periodic background.  The authors in~\cite{new3} reported an experimental study of the rogue waves on  the periodic background.

As discussed in Section IV, some elliptic-wave states are modulational
unstable. Because MI is a mechanism which generates RW patterns, we here aim
to demonstrate such an outcome of the MI-driven evolution on top of the
dnoidal-wave background.   For this purpose, we simulated the evolution of the
dnoidal-wave solutions taken as the initial condition, perturbed by a random
noise of relative strength $5\%$. The original dnoidal structure propagates
at a constant speed in the spatiotemporal contour plot, while the random
noise remains originally invisible. However, eventually, the noisy
background develops apparent chaotic dynamics. The emergence of multiple
isolated peaks at random positions, which are construed as RWs, is displayed
in Figs.~\ref{fig3} and~\ref{fig4}.  In particular, the effect of $\beta$ on the
generation of the first vector RW on top of the dnoidal-wave background are summarized in Table II.
The time of the emergence of the first RW component $S$ ($L$) is denoted as $t_{S0}$ ($t_{L0}$).
The increase of $\beta$ leads to the later appearance of the first vector RW
on top of the dnoidal-wave background, and
an increase of its amplitude.

Now we study the effect of noise in triggering RWs by decreasing the amplitude of the input noise.  From Figs.~\ref{add1} and~\ref{add2}, we find that decrease of the noise's amplitude delays the emergence of the RW patterns.

\begin{figure}[tbp]
\centering
\includegraphics[height=100pt,width=120pt]{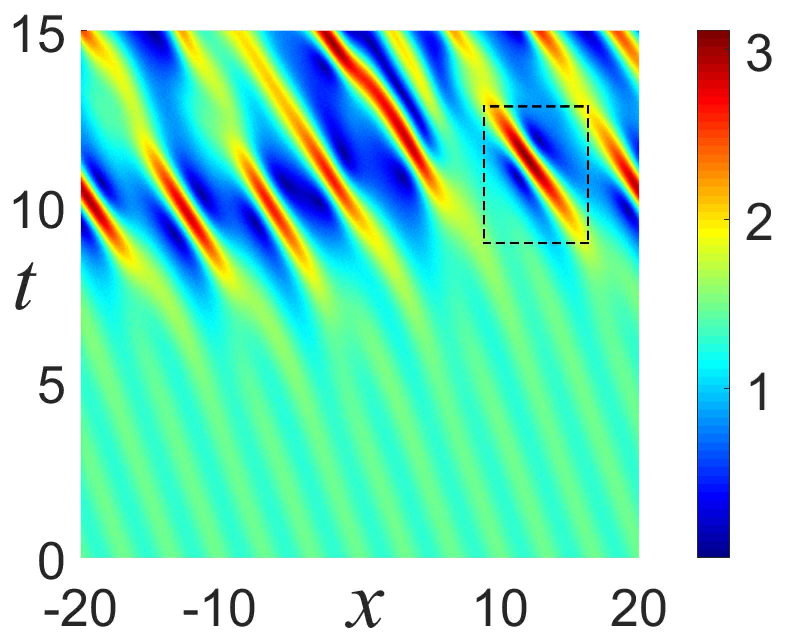} %
\includegraphics[height=100pt,width=120pt]{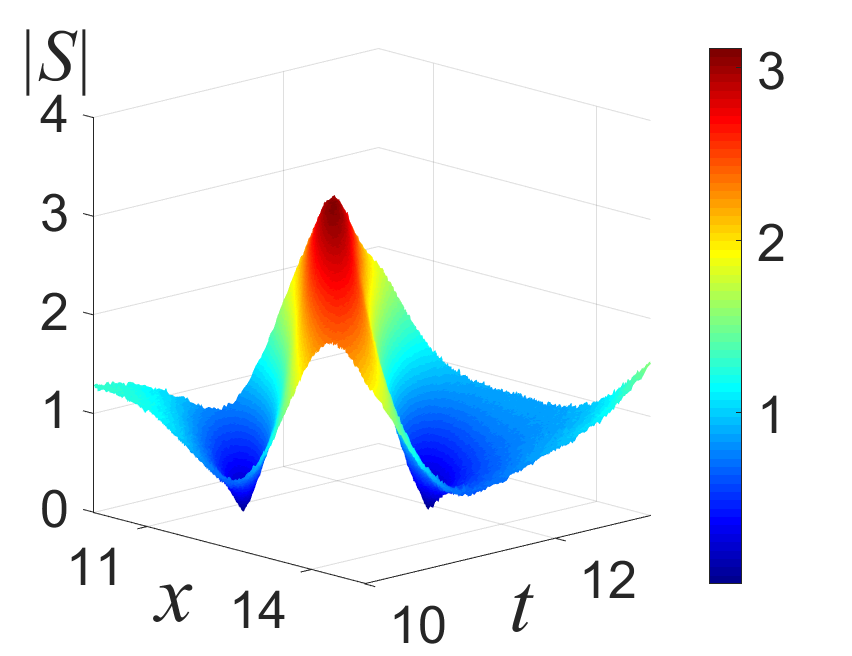} \newline
\caption{The numerically simulated formation of RWs, arising from the MI of
the dnoidal-wave background of the $S$-component of the LWSW-resonance
system with $m=1$, $\protect\omega =1$, $c=-0.9$, $k=0.5$, $\protect\beta %
=0.2$ and $\protect\sigma =-1$. A particular RW is isolated by the
surrounding box. The right panel displays the three-dimensional zoom of this
RW.}
\label{fig3}
\end{figure}

\begin{table*}[t!]
\renewcommand\arraystretch{1}
\tabcolsep 0pt \caption{The effect of $\beta$ on the first generation of vector RWs on the plane wave background}
\vspace{-0.4cm}
\label{tab-scfc}
\begin{center}
\def\temptablewidth{0.8\textwidth}
{\rule{\temptablewidth}{1.5pt}}
\begin{tabularx}{\temptablewidth}{@{\extracolsep{\fill}}p{3.2cm}cccccc}
$\beta$& 0 & 0.2 &0.4 & 0.6 &0.8\\\hline
$t_{S0}$ & 12 &15& 26&35& 65\\\hline
$max\{|S|\}$ & 2.3346 & 2.2087&2.1946 &1.9657&1.7565\\\hline
$t_{L0}$ & 12 &15& 23&35& 80\\\hline
$max\{L\}$ & 3.3443 & 3.2323 &2.9405 &2.5478 &2.4723
\vspace{0.0cm}
\end{tabularx}
{\rule{\temptablewidth}{1.5pt}}
\end{center}
\end{table*}

\begin{figure}[tbp]
\centering
\includegraphics[height=100pt,width=120pt]{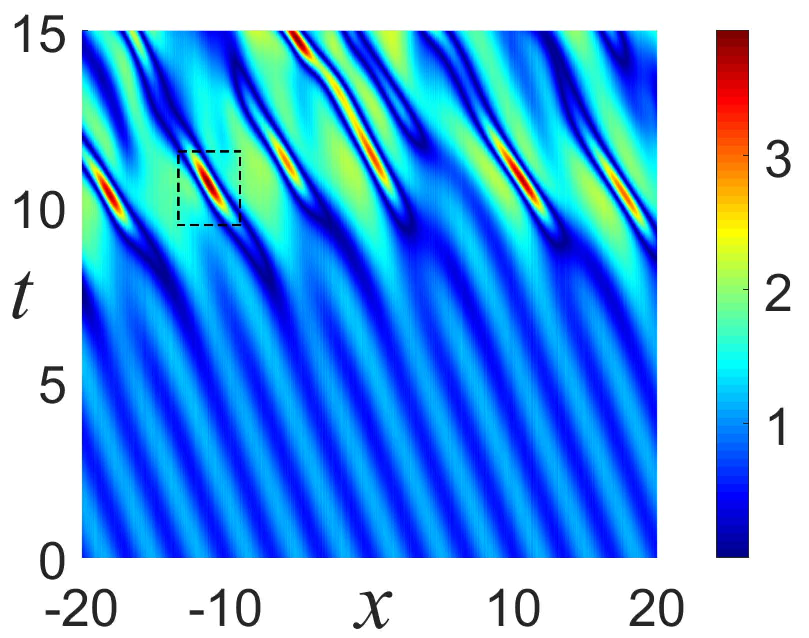} %
\includegraphics[height=100pt,width=120pt]{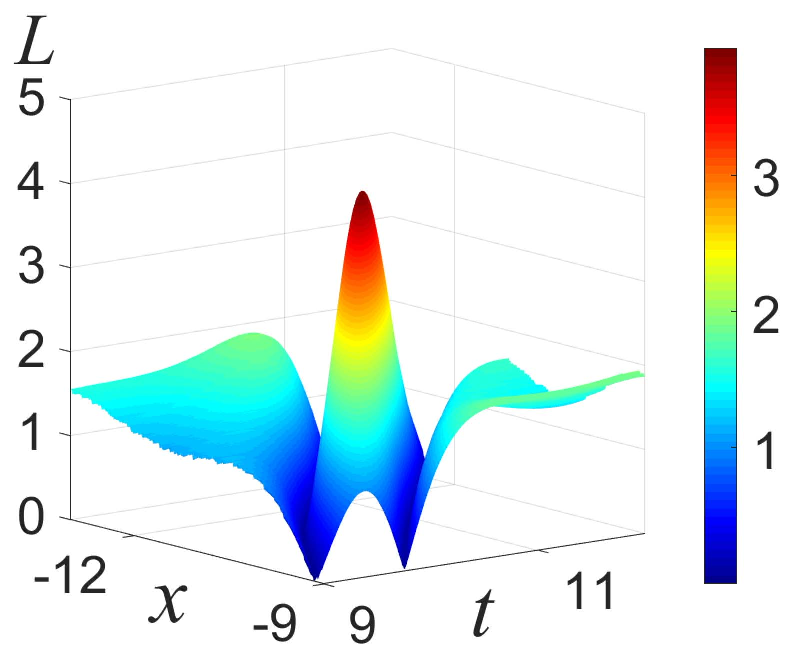} \newline
\caption{The numerically simulated formation of RWs, arising from the MI of
the dnoidal-wave background of the $L$-component of the LWSW-resonance
system with $m=1$, $\protect\omega =1$, $c=-0.9$, $k=0.5$, $\protect\beta %
=0.2$ and $\protect\sigma =-1$. A particular RW is isolated by the
surrounding box. The right panel displays the three-dimensional zoom of this
RW.}
\label{fig4}
\end{figure}

\begin{table*}[t!]
\renewcommand\arraystretch{1}
\tabcolsep 0pt \caption{The effect of $\beta$ on the first generation of vector RWs on the dnoidal wave background}
\vspace{-0.4cm}
\label{tab-scfc}
\begin{center}
\def\temptablewidth{0.8\textwidth}
{\rule{\temptablewidth}{1.5pt}}
\begin{tabularx}{\temptablewidth}{@{\extracolsep{\fill}}p{3.2cm}cccccc}
$\beta$& 0 & 0.1 &0.2 & 0.3 &0.4\\\hline
$t_{S0}$ & 10 &11& 12&14& 18\\\hline
$max\{|S|\}$ & 2.9441 & 3.0603 & 3.0803 &3.1768&3.2075\\\hline
$t_{L0}$ & 9 &10& 11&13& 15\\\hline
$max\{L\}$ & 3.8473 & 4.0363 & 4.0885 & 4.6025 & 4.6553
\vspace{0.0cm}
\end{tabularx}
{\rule{\temptablewidth}{1.5pt}}
\end{center}
\end{table*}

\begin{figure}[tbp]
\centering
\includegraphics[height=100pt,width=120pt]{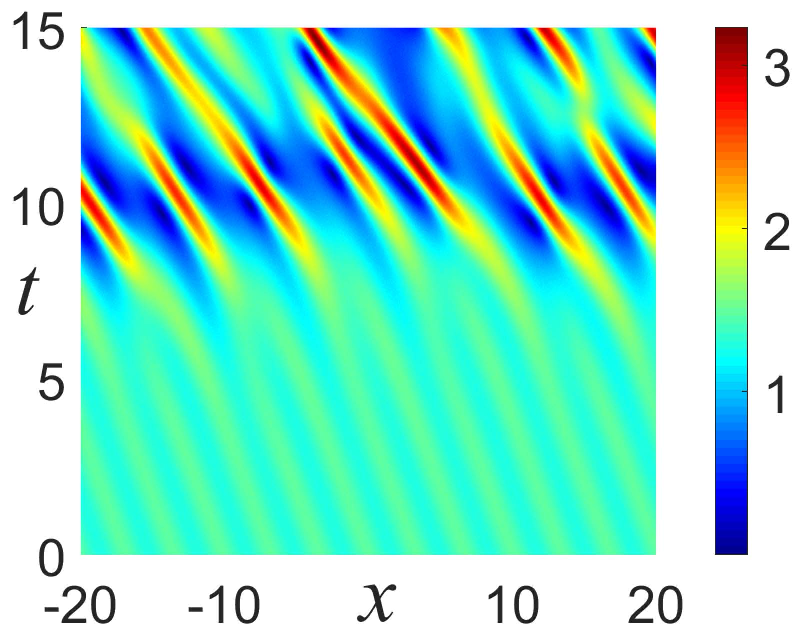} %
\includegraphics[height=100pt,width=120pt]{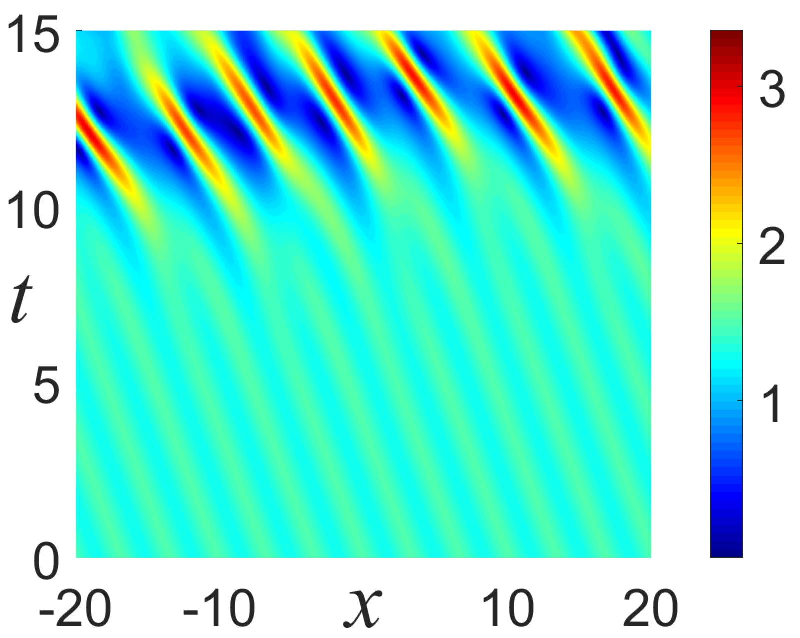} \newline
\caption{The chaotic field of $S$-component with the
dnoidal-wave solutions as the initial condition ($m=1$, $\protect\omega =1$, $c=-0.9$, $k=0.5$, $\protect\beta %
=0.2$, $\protect\sigma =-1$), perturbed by the random
noise of relative strength $3\%$ (left) and  $1\%$ (right).}
\label{add1}
\end{figure}

\begin{figure}[tbp]
\centering
\includegraphics[height=100pt,width=120pt]{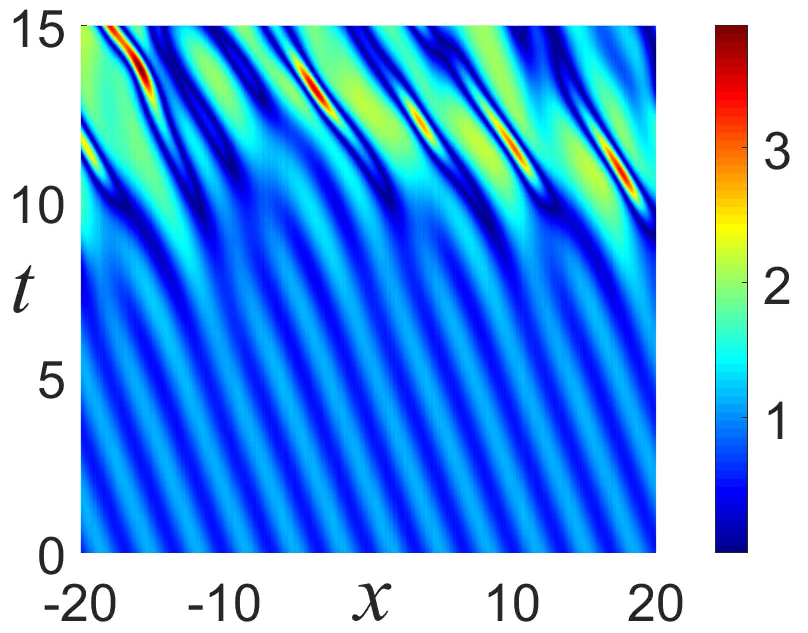} %
\includegraphics[height=100pt,width=120pt]{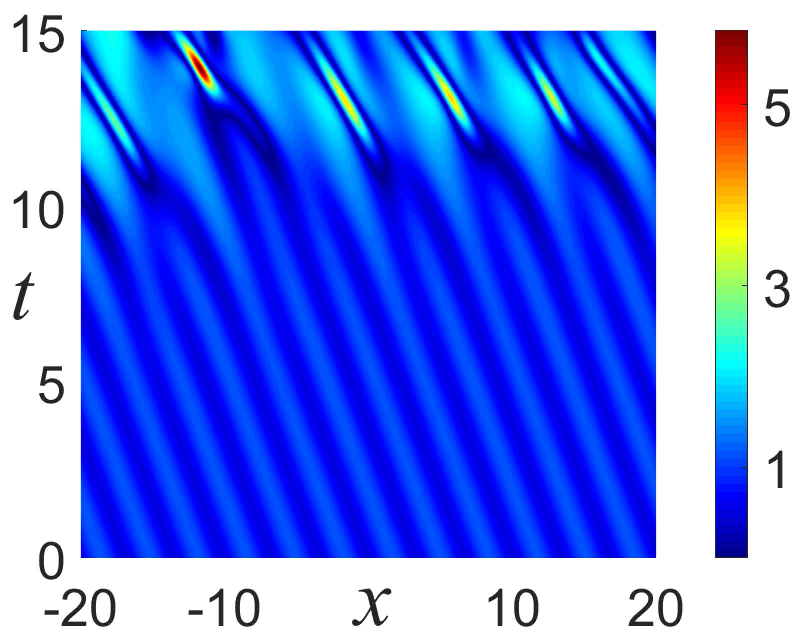} \newline
\caption{The chaotic field of $L$-component with the
dnoidal-wave solutions as the initial condition ($m=1$, $\protect\omega =1$, $c=-0.9$, $k=0.5$, $\protect\beta %
=0.2$, $\protect\sigma =-1$), perturbed by the random
noise of relative strength $3\%$ (left) and  $1\%$ (right).}
\label{add2}
\end{figure}

\section{Conclusion}

In this work, we have reported results of the systematic numerical
instability of the instabilities and RWs (rogue waves) arising from the
long-wave-short-wave (LWSW) resonance in the non-integrable system. We have
investigated the stability and instability of elliptic-waves states against
subharmonic perturbations, whose period is a multiple of the period of the
underlying elliptic-function waves. The main conclusions are:

(1) The analysis has revealed the MI (modulational instability)of the
dnoidal waves, which are stable against superharmonic perturbations. Varying
parameters of the dnoidal waves, we have displayed that a spectrally
unstable states transform into spectrally stable ones via the Hamiltonian
Hope bifurcation.

(2) We have identified the dominant instability scenarios driven by the
competing MI and bubble-like instability mechanisms for the snoidal waves.
For the cnoidal waves, we have found three different scenarios of the MI.

(3) We have systematically simulated the emergence of the RWs in the
LWSW-resonance system on top of the CW and elliptic-wave backgrounds,
initiated by random perturbations.

Since system~(\ref{lwswm}) is not integrable, we can't construct the doubly-localized Peregrine solutions analytically.  The lack of integrability makes it necessary to develop a detailed numerical analysis aimed at the search for Peregrine-like solutions in system~(\ref{lwswm}), which should be a subject of a separate work.

\section*{Acknowledgments}

This work has been supported by the Fundamental Research Funds of the
Central Universities of China (No. 230201606500048). The work of B.A.M. is
supported, in part, by the Israel Science Foundation (Grant No. 1695/22).

\end{subequations}

\end{document}